\documentclass[12pt]{article}
\usepackage{bezier}
\usepackage{amssymb}
\usepackage{epsfig}

\newtheorem{lemma}{Lemma}

\newtheorem{proposition}{Proposition}
\newtheorem{definition}{Definition}

\newcommand{\nc}{\newcommand}
\nc{\be}{\begin{equation}}
\nc{\ee}{\end{equation}}
\nc{\bea}{\begin{eqnarray}}
\nc{\eea}{\end{eqnarray}}

\nc{\eqn}[1]{{(\ref{#1})}}
\nc{\cA}{{\cal A}}
\nc{\cB}{{\cal B}}
\nc{\cC}{{\cal C}}
\nc{\cD}{{\cal D}}
\nc{\cE}{{\cal E}}
\nc{\cF}{{\cal F}}
\nc{\cG}{{\cal G}}
\nc{\cH}{{\cal H}}
\nc{\cI}{{\cal I}}
\nc{\cJ}{{\cal J}}
\nc{\cK}{{\cal K}}
\nc{\cL}{{\cal L}}
\nc{\cM}{{\cal M}}
\nc{\cN}{{\cal N}}
\nc{\cO}{{\cal O}}
\nc{\cP}{{\cal P}}
\nc{\cQ}{{\cal Q}}
\nc{\cR}{{\cal R}}
\nc{\cS}{{\cal S}}
\nc{\cT}{{\cal T}}
\nc{\cU}{{\cal U}}
\nc{\cV}{{\cal V}}
\nc{\cW}{{\cal W}}
\nc{\cX}{{\cal X}}
\nc{\cY}{{\cal Y}}
\nc{\cZ}{{\cal Z}}
\nc{\simo}[1]{{\stackrel{#1}{\simeq}}}
\nc{\geqo}[1]{{\stackrel{#1}{\geq}}}
\nc{\geo}[1]{{\stackrel{#1}{>}}}
\nc{\guo}[1]{{\stackrel{#1}{\succ}}}

\nc{\rbo}{\raisebox}
\nc{\RR} {\rangle \! \rangle}
\nc{\LL} {\langle \! \langle}
\nc{\rmi}[1]{{\mbox{\small #1}}}
\nc{\eq}{eq.~}
\nc{\nr}[1]{(\ref{#1})}
\nc{\ul}{\underline}
\nc{\mc}{\multicolumn}
\nc{\todo}[1]{\par\noindent{\bf $\rightarrow$ #1}}

\nc{\cu}{{\cal u}}

\title{
  \begin{flushright} {\small $\begin{array}{ l } \mbox{HD--THEP--98--06} \\
    \mbox{WUB--98--03} \end{array} $}
 \end{flushright}
The Monotony Criterion for a \\
Finite Size Scaling Analysis of  \\
Phase Transitions}

\author{Hildegard~Meyer-Ortmanns\thanks{E-mail address:
ortmanns@theorie.physik.uni-wuppertal.de}
         \\ \\Institut
        f\"ur Theoretische Physik,\\
        Bergische Universit\"at Wuppertal, \\
        Gau\ss strasse 20, \\
        D-42097 Wuppertal, Germany
  \\ \\ and \\ \\
 Thomas~Reisz\thanks{Supported by a Heisenberg Fellowship,
      E-mail address: reisz@thphys.uni-heidelberg.de}
         \\ \\Institut
        f\"ur Theoretische Physik,\\
        Universit\"at Heidelberg, \\
        Philosophenweg 16, \\
        D-69120 Heidelberg, Germany}

\begin{document}

\maketitle

\begin{abstract}
We propose a new criterion to analyse the order of phase transitions
within a finite size scaling analysis.
It refers to response functions like order parameter susceptibilities
and the specific heat and states different monotony behaviour in volume
for first and second order transitions close to the transition point.
The criterion applies to analytical and numerical studies of phase
diagrams including tricritical behaviour.

\end{abstract}

%
%
%
\section{\label{mon.intro} Introduction}
%
Finite size scaling analysis is particularly useful in numerical
simulations to distinguish first from second order phase transitions
by their different onset
in a large but finite volume. Usually it is the specific scaling
behaviour of the peak in a susceptibility, the shift in the critical
coupling/temperature and the width of the critical region that
anticipates the nature of the transition in the thermodynamic 
limit \cite{barber}.
\noindent

%
%
\noindent
In this letter we formulate a new criterion in a finite volume to
distinguish
first from second order phase transitions. 
For a certain interval of the scaling region response
functions  with 
a nonanalytic behaviour in the infinite volume limit show different
monotony behaviour for 1st and 2nd order transitions.
Examples for such functions are the specific heat and order parameter
susceptibilities. They are
increasing in volume in a certain neighbourhood of $T_c$ for 2nd order
transitions, and decreasing for 1st order transitions for some
range in the scaling region, which will be specified below.

\begin{figure}[htb]
\caption{\label{monfig1} $(t,v)$-plane in the vicinity of a phase transition at
$t=v=0$. $t$ denotes the scaling field
$t=(T-T_c(L))/T_c(L)$, $v$ is inverse to some power of the volume $L^x$
with $x>0$.
Let us consider the susceptibility as a function of $t$ and $v$.
Then, for a 1st~order transition, $\chi(t,v_1)>\chi(t,v_2)$,
whereas for a 2nd~order transition
$\chi(t,v_1)<\chi(t,v_2)$. The shaded part is the crossover region
between different asymptotic behaviour as $(t,v)\to 0$.}

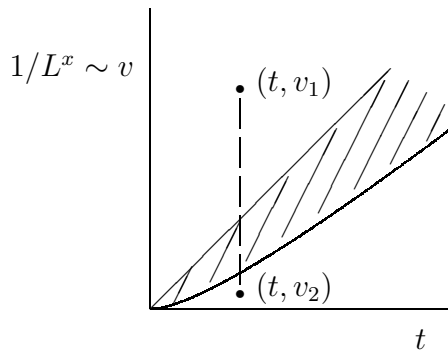
\begin{center}
\setlength{\unitlength}{0.8cm}
\begin{picture}(10.0,5.0)


\put(2.0,0.0){
\setlength{\unitlength}{0.8cm}
\begin{picture}(10.0,5.0)

\put(0.0,0.0){\line(1,0){5.0}}
\put(-1.8,4.0){\makebox(1.0,0){$1/L^x\sim v$}}
\put(0.0,0.0){\line(0,1){5.0}}
\put(4.0,-0.5){\makebox(1.0,0){$t$}}

\qbezier(0.0,0.0)(1.0,0.0)(5.0,3.0)
\put(0.0,0.0){\line(1,1){4.0}}

\put(1.5,3.65){\circle*{0.12}}
\put(1.9,3.75){\makebox(1.0,0){$(t,v_1)$}}
\put(1.5,0.25){\circle*{0.12}}
\put(1.9,0.35){\makebox(1.0,0){$(t,v_2)$}}
\put(1.5,0.4){\line(0,1){0.5}}
\put(1.5,1.05){\line(0,1){0.5}}
\put(1.5,1.7){\line(0,1){0.5}}
\put(1.5,2.35){\line(0,1){0.5}}
\put(1.5,3.0){\line(0,1){0.5}}
\put(0.4,0.1){\line(1,2){0.3}}
\put(1.0,0.45){\line(1,2){0.5}}
\put(1.6,0.8){\line(1,2){0.7}}
\put(2.2,1.2){\line(1,2){0.9}}
\put(2.8,1.6){\line(1,2){1.1}}
\put(3.4,2.0){\line(1,2){0.9}}
\put(4.0,2.4){\line(1,2){0.6}}
\put(4.6,2.8){\line(1,2){0.3}}

\end{picture}
}

\end{picture}
\end{center}

\end{figure}


\noindent
For definiteness we fix the notation in terms of an order parameter
susceptibility $\chi(T,L)$, considered as a function of the temperature
$T$ and the spatial size parameter $L$, so that a symmetric
$D$-dimensional volume is of size $L^D$.
By $T_c(L)$ we denote the location of the maximum
of $\chi(\cdot , L)$ in volume $L^D$ associated with the
phase transition at $T_c \equiv T_c(\infty)$.
In the infinite volume
\be \label{monintro.2}
  \chi (t + T_c(\infty), L=\infty ) \; < \; \infty 
   \quad \mbox{as} \; t\to 0
\ee
for a 1st order transition at $T_c$,
with a possible discontinuity at $t=0$
in the associated order parameter, $t$ denoting the scaling field
$t=(T-T_c(L))/T_c(L)$,
whereas
\be \label{monintro.3}
  \chi (t + T_c(\infty), L=\infty) \simeq \cA |t|^{-\gamma}
   \quad \mbox{as} \; t\to 0
\ee
for a 2nd order transition with critical exponent $\gamma>0$.
On the other hand, in the opposite order of limits, $\chi(T_c(L), L)$ 
diverges in both cases as
$L$ approaches infinity. Stated differently,
at $T_c(L)$, in the limit $L\to\infty$ $\chi$ has a
$\delta$-function singularity or a power law type of
singularity for a 1st or 2nd order transition, respectively.
It is this difference
that is responsible for
the different monotony properties for $t\not=0$ in the finite volume.

The essential statement of the monotony criterion is the following
(details will be given below):
For a sufficiently large size $L_s < \infty$,
there is always a size $L_l$ with
$L_l>L_s$ and such that (cf. Fig.~\ref{monfig1})
\bea \label{monintro.4}
  \chi(\delta + T_c(L), L\geq L_l ) < \chi(\delta + T_c(L_s), L_s)
   \quad & & \mbox{for 1st order} , \\ \label{monintro.5}
    \chi(\delta + T_c(L), L\geq L_l ) > \chi(\delta + T_c(L_s), L_s)
   \quad & & \mbox{for 2nd order} , 
\eea
that is, $\chi(\delta + T_c(L),L)$ is decreasing or increasing in volume
for a 1st or 2nd order phase transition, respectively.
The susceptibility is measured at fixed distance $\delta$ to the 
volume dependent maximum, constraint by
\be \label{monintro.45}
  c_1 \sigma(L)^{\frac{1}{1+\epsilon}} < |\delta| < 
  c_2 \sigma(L_s).
\ee
Here $\sigma(L)=L^{-D}$ or $\sigma(L)=L^{-1/\nu}$ for 1st or 2nd
order transition, respectively, with $\nu$ being the critical 
exponent of the correlation
length, $\xi\sim |T-T_c|^{-\nu}$ as $T\to T_c$.
Hence $\sigma(L)$ is the width of the critical region
in the volume $L^D$,
$c_1$, $c_2$ and $\epsilon$ are positive constants,
typically $\epsilon =1$.
Beyond the general constraint that both linear sizes $L_s$ and $L_l$ 
have to be  sufficiently large, in addition 
$L_l$ has to be sufficiently larger than $L_s$, so that
$\sigma(L_l)$ is considerably smaller than $\sigma(L_s)$.
The monotony behaviour of $\chi$
does {\sl not} refer to values of $L_l$ close to $L_s$.
In terms of Fig.~\ref{monfig1}, $L_s$ and $L_l$ are separated
by the shaded area, for which we do not make any predictions
on the behaviour of $\chi$.
 
In section 2 we make these statements more precise. 
We specify the 
conditions under which the criterion holds and give the proof.
As it turns out, the criterion applies whenever the standard
conjectures on finite size scaling behaviour hold. 
Rigorous proofs of these standard conjectures are missing in general,
neither do we attempt to give such proofs
in the treatment below. 
In section 3 we conclude with some remarks on testing the criterion in
analytical and numerical calculations.

%
%
\section{\label{mon.lemmas} Statements of the criterion}

Let
$t$ denote the scaling field, i.e. $t=(T-T_c(L))/T_c(L)$.
$T_c(L)$ locates the maximum of the susceptibility in
the volume $L^D$.
Furthermore we set $v=L^{-m}$ with some $m>0$.
The infinite volume limit is obtained
as $v\to 0$ from above. 

The transition region is given by
small $t$ and $v$.
Let
\be \label{mon.h2}
   H^2 := \{ (t,v)\in{\bf R}^2 \; \vert
     \; v\geq 0 \}
\ee
denote a half plane, $\cU\subseteq H^2$ the intersection of an open 
neighbourhood
of $0\in{\bf R}^2$ with $H^2$ and $\cU^* = \cU\setminus\{0\}$.

%
%

\subsection{First order transitions}

The typical behaviour of a susceptibility close to a first order
transition is summarized by the following definition
(with $v=L^{-D}$).
As usual, $C(\cU^*)$ denotes the set of real valued continuous
functions in $\cU^*$.


\begin{definition}\label{mondef1}
Let $\omega >0$. $\Psi_1^\omega(\cU)$ is the set of functions
$\chi\in C(\cU^*)$
with the following properties.
\begin{enumerate}
\item
$\chi\in C^1(\cU^*\setminus\{({\bf R},0)\})$,
that is, $\chi$ is continuously
differentiable for $v\not=0$.
\item
There exists $\nu(t)>0$ such that for all
$v < |t|/\nu(t)$
\[
    | \; \chi(t,v) \; | \leq \omega .
\]
\item
With appropriate positive constants $c,K_1,K_2$ and $\epsilon$
we have in $\cU^*$ for $v\not= 0$
\bea
  | \; \chi (0,v) - \frac{c}{v} \; | & < & \frac{K_1}{v^{1-\epsilon}} ,
  \nonumber \\
  | \; \frac{\partial}{\partial t} \chi(t,v) \; |
  & < & \frac{K_2}{v^2} . \nonumber
\eea
\end{enumerate}
\end{definition}

It follows already from this definition that $\nu(t)$ is bounded from below
by some positive constant.
This is part of the following proposition.
The main statement is the monotony behaviour of a function that
obeys Definition \ref{mondef1}.


\begin{proposition}\label{monlemma1}
Let $\omega>0$, $\chi \in \Psi_1^\omega(\cU)$.
There exist positive $\delta$, $\epsilon$ such that
the following statements hold.
\begin{enumerate}
\item
For all $t$ with $|t|<\epsilon\delta$,
\be \label{nut}
   \nu(t) > \epsilon .
\ee
\item
For all $(t,v),(t,w)\in\cU^*$ with
$\nu(t)w < |t| < \epsilon v$ and $v<\delta$,
we have $w<v$ and
\be
   \chi(t,v) \; > \; | \chi(t,w)| .
\ee \label{mondecr}
\end{enumerate}
\end{proposition}

The proposition includes the case $w=0$ of infinite volume.
In this case, in the critical region $|t|/v < \epsilon$, the
susceptibility $\chi$ is always larger in the finite volume
than in the infinite one.

{\sl Proof:}
Let $\chi\in\Psi_1^\omega(\cU)$, $\omega>0$.

$\alpha$.
There are $\epsilon,\delta>0$ such that for all $t,v$ with
$v<\delta$, $|t|<\epsilon v$
\be \label{mon.twoomega}
   \chi(t,v) \geq 2\omega.
\ee
{} For, 
\[
   \chi(t,v) = \chi(0,v) + t \int_0^1 ds
   \left. \frac{\partial}{\partial\eta} \chi(\eta,v)
   \right|_{\eta = st},
\]
and there are $c_0,K>0$ so that
\[
   \chi(0,v) > \frac{c_0}{v},
\]
\[
   \left| \frac{\partial}{\partial\eta} \chi(\eta,v)
   \right| < \frac{K}{v^2}
\]
in $\cU^*$. Hence, with $\epsilon = c_0 /(2K)$,
\[
   \chi(t,v) \geq \frac{c_0}{v} - \frac{|t| K}{v^2}
   > \frac{c_0 - \epsilon K}{v} = \frac{c_0}{2 v}.
\]
Choosing $\delta = c_0/(4\omega)$, we get for $v < \delta$
the lower bound (\ref{mon.twoomega}) on $\chi(t,v)$.

$\beta$.
By definition of $\nu(t)$, we have whenever $\nu(t) w < |t|$
the upper bound
\be \label{strictdecr}
   | \chi(t,w) | \leq \omega .
\ee

$\gamma$.
Summarizing $\alpha$ and $\beta$,
it follows that for all $w,v$ with
$\nu(t) w < |t| < \epsilon v$, $v < \delta$, the strict bound
\[
   |\chi(t,w)| \; < \; \chi(t,v)
\]
holds. Furthermore, (\ref{strictdecr}) implies that
$|\chi(t,w^{\,\prime})|\leq\omega$ for all
$0\leq w^{\,\prime}\leq\omega$.
Hence it must hold that $w<v$.
It follows that
\[
  \frac{|t|}{\nu(t)} \leq \frac{|t|}{\epsilon}
\]
and hence $\nu(t)>\epsilon$ for $|t|<\epsilon\delta$.
$\qquad\square$
\vskip2pt
The remainder of this section is devoted to the function class
$\Psi_1^\omega(\cU)$.
As an example of such a function we consider the following typical
representation of the magnetic susceptibility in the volume $L^D$
close to a first order phase transition,
\be \label{mon.chi2}
   \chi_2(T ,L) = c_2 L^D \; \exp{(-f_2 L^{2D}
    (T - T_c(L))^2 )}
    + \eta_2(T,L).
\ee
Here $c_2, f_2$ are positive real constants
and $\eta_2(\cdot,L)$ is analytic for $L<\infty$,
locally uniformly convergent as $L\nearrow\infty$
(so that $\eta_2(\cdot,\infty)$ is analytic).
Furthermore, $T_c(L) = T_c - d_2 L^{-m} + o(L^{-m})$, with $m\geq 0$.

We can view $\chi_2$ as a function of some $\Psi_1^\omega$ in
essentially two ways.
The first one corresponds to the point of view given in the
introduction, in which the scaling field $t$ measures the distance
of the temperature $T$ to the volume dependent location of 
the peak $T_c(L)$.
It does not refer to a particular volume dependence of $T_c(L)$
itself.
That is, $m$ does not need further specification except that $m\geq 0$.

We set $v = L^{-D}$, $t = (T-T_c(L))/T_c(L)$ and
\be \label{mon.exa1}
  \chi(t,v) = \chi_2(T,L) = c v^{-1} \;
  \exp{(- f(v) v^{-2} t^2)} \; + \eta(t,v),
\ee
where $\eta(t,v)\equiv \eta_2(T,L)$ and
$f(v) = f_2 T_c(L)^2$, $c=c_2$.


\begin{lemma}\label{monlemma3}
$\chi \in \Psi_1^\omega(\cU)$ for some $\omega>0$ and sufficiently
small $\cU$.
$\nu(t)$ of Def.~\ref{mondef1} satisfies
$\nu(t)^{-1} \to 0$ as $|t|\to 0$.
\end{lemma}

{\sl Proof:} 
For $\alpha,\beta>0$ and $x\in{\bf R}$ the function
\[
  f_{\alpha,\beta}(x) \; = \; x^\alpha \;
  \exp{(-\beta x^2)}
\]
attains its maximum at $x^2=\alpha/(2\beta)$, is positive and
monotone decreasing for $x>\left(\alpha/(2\beta)\right)^{1/2}$,
and satisfies on ${\bf R}$ the bound
\[
  | f_{\alpha,\beta}(x) | \; \leq \;
  \left(\frac{\alpha}{2\beta e}\right)^{\frac{\alpha}{2}}.
\]
We prove that $\chi(t,v)$ match the criteria of
Definition \ref{mondef1} to belong to some $\Psi_1^\omega(\cU)$.
There is $\delta>0$ such that for all $v<\delta$ we have
$\widehat{f}\leq f(v) \leq \overline{f}$ for some
$\widehat{f}, \overline{f}>0$.
We choose $\cU$ such that $(t,v)\in\cU$ implies that
$v <\delta$ and $|t|<\widehat{f}^{1/2}$ and put
\[
  K_\eta \; = \; \max_{(t,v)\in\cU}
  \left( |\eta(t,v)|, |\partial_t\eta(t,v)| \right).
\]
Condition 1.~is obviously satisfied.
To show 2.~we notice that for
\[
  \frac{|t|}{v} \; > \; \nu(t) \; > \; 
  \left(\frac{1}{2 \widehat{f}}\right)^{\frac{1}{2}}
\]
one obtains
\[
   |\chi(t,v)| \; < \; 
   \left| \frac{c}{|t|} \, \frac{|t|}{v} \;
    \exp{\left(- f(v) \left(\frac{|t|}{v}\right)^2 \right)}
   \; + \eta \right|
   \; \leq \;
   c \, \frac{\nu(t)}{|t|} \; \exp{(-f(v) \nu(t)^2)}
   + \left| \eta \right|.
\]
We choose $\nu(t)=|t|^{-1}$ and obtain with
$\omega = c/(\widehat{f}e)+K_\eta$
\[
  | \chi(t,v) | \; \leq \; \omega .
\]
To verify the 3rd condition, we compute
\[ 
   | \chi(0,v) - c v^{-1} | = |\eta(t,v)| \leq K_\eta,
\]
and notice that there is $K>0$ such that
\bea \nonumber
   \left| \frac{\partial}{\partial t} \chi(t,v)
   \right|  & \leq &
   \frac{2 |t|}{v^2} \, f \, |\chi| + | \partial_t\eta(t,v) | 
   \\ \nonumber & \leq &
   \frac{2\overline{f}}{v^2}
   \left\lbrack c 
      \left(\frac{1}{2\widehat{f} e}\right)^{\frac{1}{2}}
      + \widehat{f}^\frac{1}{2} K_\eta
   \right\rbrack
   \; + \; K_\eta \\ \nonumber
   & \leq &
   \frac{K}{v^2}.
\eea
Hence, $\chi\in\Psi_1^\omega(\cU)$.
$\qquad\square$
\vskip2pt
The next lemma states that the susceptibility $\chi_2$
matches the criteria of Definition \ref{mondef1}
also if the scaling field $t$ measures the distance of $T$
from the phase transition point, i.e.~to
$T_c(L=\infty)$. The only additional constraint
then is that the peak $T_c(L)$ of $\chi_2$ converges
sufficiently fast to $T_c(\infty)$, namely that
\be \label{running.peak}
  \left| T_c(L) - T_c(\infty) \right|
  \; \leq \; \frac{c}{L^m}
 \qquad\mbox{with}\; m\geq D.
\ee
We set
$v = L^{-D}$, $t = (T-T_c(\infty))/T_c(\infty)$ and
\be \label{mon.exa2}
  \chi(t,v) = \chi_2(T,L) = c v^{-1} \;
  \exp{(- f v^{-2} (t+d v^n)^2)} \; + \eta(t,v),
\ee
where now $\eta(t,v)\equiv \eta_2(T,L)$ and
$f=f_2 T_c(\infty)^2$, $d=d_2/T_c(\infty)$,
and $n = m/D$.
Similarly as above, we get


\begin{lemma}\label{monlemma4}
Let $n\geq 1$. Then
$\chi \in \Psi_1^\omega(\cU)$ for some $\omega>0$ and sufficiently
small $\cU$.
$\nu(t)$ of Def.~\ref{mondef1} satisfies
$\nu(t)^{-1} \to 0$ as $|t|\to 0$.
\end{lemma}

{\sl Proof:} 
For $a,b,\epsilon\in{\bf R}$, $\epsilon\not= 0$,
\[
  ( a + b )^2 \geq a^2 \left( 1 - \frac{1}{\epsilon^2} \right)
    + b^2 \left( 1 - \epsilon^2 \right).
\]
We choose $\delta,\eta>0$ sufficiently small
such that the following estimates on $\chi$ are true for all
$t,v$ with $|t|<\eta$, $0<v<\delta$, identifying $\cU$.

$\chi$ obviously satisfies condition 1. of Def.~\ref{mondef1}.
To show 2.~we estimate
\bea \nonumber
   |\chi(t,v)| &\leq& c v^{-1} \; 
    \exp{(- f v^{-2} (\frac{1}{2} t^2 - (d v^n)^2) )} \; 
    + \left| \eta \right|  \\
   &\leq&  2 c v^{-1} \; \exp{(-\frac{f}{2} v^{-2} t^2)} 
   + \left| \eta \right|.
\eea
Proceeding as in the proof of the previous lemma
we obtain condition 2,~with appropriate $\omega$.
To realize the 3rd condition, we first notice that
\[ 
   | \chi(0,v) - c v^{-1} | \leq c v^{-1} 
   \left| e^{-fd^2 v^{2(n-1)}} - 1 \right| + |\eta|
   \leq \frac{K}{v^{1-\epsilon}}
\]
for some positive constant $K$, with $\epsilon = \min{(1, 2(n-1))}\geq 0$.
Finally there is $K^{\,\prime}>0$ such that
\bea \nonumber
     \left| \frac{\partial}{\partial t} \chi(t,v)
   \right| &=&
   2 f v^{-2} |t+d v^n| |\chi| + | \partial_t\eta(t,v) | \\
   &\leq& 2 f c v^{-2} 
   \left(\frac{1}{2 f e}\right)^{\frac{1}{2}}
   \cdot 2 \; = \; \frac{K^\prime}{v^2}.
\eea
Hence $\chi\in\Psi_1^\omega(\cU)$.
$\qquad\square$

\vskip2pt
Finally, we state without proof that every function
$\chi:\cU^*\to{\bf R}$ of the form
\be \label{monf1o}
   \chi(t,v) \; = \; \frac{1}{v} \; f(\frac{t}{v})
   + \widetilde\chi(t,v)
\ee
belongs to $\Psi_1^\omega(\cU)$ for some $\cU$ and
with appropriate $\omega>0$,
if the following conditions are satisfied.
\begin{description}
\item{1a.} $\widetilde\chi(t,v)\in C^1(\cU^*)$,
\item{1b.} $\widetilde\chi(t,v)$ together with its (first) partial
derivatives are uniformly bounded in $\cU^*$.
\item{2a.} $f\in C^1({\bf R})$ is a nonnegative function with $f(0)>0$,
\item{2b.} $\lim_{x\to\pm\infty} |x|^{1+\epsilon} f(x) = 0$
for some $\epsilon>0$,
\item{2c.} $(d/dx)f(x)$ is uniformly bounded on ${\bf R}$.
\end{description}

Any such function has the property that it "approaches $\delta$"
locally, i.e.~there is $c>0$ so that for any test function $g(t)$
supported in a small neighbourhood of $t=0$
\[
    \lim_{v\to 0+} [\chi(t,v)] g \; = \;
    \lim_{v\to 0+}
    \int dt \; \chi(t,v) g(t) \; = \; c g(0).
\]
The example discussed above corresponds to
\[
     g(x) \; = \; (\frac{c}{\pi})^{1/2} \exp{(-cx^2)} \; , \; c>0.
\]

%
%

\subsection{Second order transitions}

In contrast to the first order case, at a second order transition
the susceptibility $\chi$ can be divergent
in the infinite volume limit
as the critical temperature is approached.
(It need not be divergent, since another second derivative
of a suitable thermodynamic potential may be divergent
instead.)
If $\chi$ is divergent with a critical exponent, it is
described by the following definition
(with $v=L^{-1/\nu}$, $\nu>0$ being the critical exponent
of the correlation length $\xi$).

\begin{definition}\label{mondef2o}
Let $\gamma >0$. $\Psi_2^\gamma(\cU)$ is the set of functions
$\chi\in C(\cU^*)$ that satisfy the following conditions.
\begin{description}
\item{1.} There are constants $\cA$, $K$, $\epsilon>0$ such that in $\cU^*$
\[
    | \; \chi(t,0) \; - \; \cA\; |t|^{-\gamma} \; |
    \leq \; K \; |t|^{-\gamma+\epsilon} .
\]
Furthermore, there exist $\nu$, $\cC>0$ so that for
$|t| > \nu\; v$,
\[
   \chi(t,v) \; \geq \; \cC \; \chi(t,0).
\]
\item{2.} There are constants $\eta$, $\cB>0$ such that for $|t|<\eta\; v$
\[
   | \; \chi(t,v) \; | \; < \; \cB \; v^{-\gamma}.
\]
\end{description}
\end{definition}

Functions that obey Definition~\ref{mondef2o}
are always increasing in volume close to the transition.

\begin{proposition}\label{monlem2o}
Let $\gamma>0$ and $\chi\in\Psi_2^\gamma(\cU)$.
There are constants $\nu$, $\epsilon>0$
such that for all $t,v,w$ with $\nu w < |t| < \epsilon v$
we have $w<v$ and
\[
   | \; \chi(t,v) \; | \; < \; \chi(t,w) .
\]
\end{proposition}

The inequality includes the case $w=0$, corresponding to
infinite volume. That is,
the susceptibility is always smaller in finite volume than in the infinite
volume as long as we are in the critical
region $|t|/v < \epsilon$.

{\sl Proof:}
Let $\chi\in\Psi_2^\gamma(\cU)$, $\gamma>0$.
There are numbers $\cC,\cD,\nu>0$ such that in $\cU^*$ for
$|t| > \nu\; w$
\[
   \chi(t,w) \; > \; \cC \; \chi(t,0) \; > \; \cD \; |t|^{-\gamma}.
\]
Furthermore, there are $\eta,\cB>0$ such that for
$|t| < \eta\; v$
\[
   | \chi(t,v) | \; < \; \cB \; v^{-\gamma}.
\]
We choose $\epsilon=\min{(\eta, (\cD/\cB)^{1/\gamma})}$
and get for
$\nu\; w < |t| < \epsilon\; v$
\[
   | \chi(t,v) | \; < \;  \cB \; v^{-\gamma}
   \; < \; \cB \left(\frac{\epsilon}{|t|}\right)^\gamma
   \; < \; \frac{\cB\epsilon^\gamma}{\cD} \; \chi(t,w)
   \; < \; \chi(t,w).
\]
This proves the lemma.
$\qquad\square$

\vskip2pt
The specific  property for a 2nd order transition
that the singular part of the
free energy density behaves as a generalized homogeneous
function implies for the susceptibility in a volume $L^D$
a typical form like
\be
   \chi_2(T ,L) = |T-T_c|^{-\gamma} \; Q( (T-T_c) L^{1/\nu} )
    + \eta(T,L),
\ee
with some $\gamma >0$.
Here $\eta(\cdot,L)$ has similar analyticity properties as in \eqn{mon.chi2}
above, $\nu>0$ is the critical exponent of the correlation length
$\xi \sim | T-T_c |^{-\nu}$.
$Q$ is continuous and behaves as
\bea 
   \lim_{x\to 0} |x|^{-\gamma} Q(x) & = & K > 0, \nonumber \\
    \lim_{x\to\pm\infty} Q(x) & = & C > 0 . \nonumber
\eea
The first equation expresses
the absence of a divergence of $\chi_2$ for finite $L$, the
second one its presence in the infinite volume case.
With $t=(T-T_c)/T_c$, $v=L^{-1/\nu}$ and
\[
   \chi(t,v) \; := \; \chi_2(T,L)
\]
it is straightforward to verify that $\chi$ belongs to
$\Psi_2^\gamma(\cU)$.

More generally, every function $\chi:\cU^*\to{\bf R}$ of the form
\be
   \chi(t,v) \; = \; \frac{1}{v^\gamma} \; f(\frac{t}{v})
   + \widetilde\chi(t,v)
\ee
with $\gamma >0$
belongs to some $\Psi_2^\gamma(\cU)$, if the following conditions
are satisfied:
\begin{description}
\item{1.} $\widetilde\chi(t,v)\in C^1(\cU^*)$.
\item{2a.} $f\in C^1({\bf R})$, and $f(0)>0$,
\item{2b.} $\lim_{x\to\pm\infty} |x|^{\gamma} f(x) = C$
for some finite $C>0$.
\end{description}

Compared to the first order case \eqn{monf1o} the
essential difference comes from property (2b).

As an example,
\[
    \chi(t,v) \; = \; (t^2+v^2)^{-(m/2)} \;, \; m>0,
\]
belongs to the class
$\Psi_2^{m}$.

%
\section{\label{monte} Applications}
%
The monotony criterion has been applied in the
framework of convergent series expansions of $O(N)$ symmetric 
spin models on the lattice, performed to the 20th order
in the hopping parameter (corresponding to the inverse temperature)
both in the infinite and finite volume \cite{jsp}.
By means of the series representation it is hard to compute the
scaling behaviour of a susceptibility $\chi(T,L)$ at the
maximum $T_c(L)$, since $T_c(L)$ as well as the transition at
$T_c(\infty)$ are
extrapolated as the convergence radius of the series.
The monotony criterion circumvents this difficulty in that it
allows us to determine the order of the transition by the
different $L$-dependence of $\chi(T,L)$ already close to $T_c(L)$,
that is by an
increase in volume for a second order transition, and a decrease for
a first order transition.

In series representations the larger volume $L_l^D$ is conveniently
chosen to be infinite so that (\ref{monintro.4})-(\ref{monintro.45}),
with $L=\infty$ and $\sigma(L)=0$,
give strong criteria in the whole scaling region.
In Monte Carlo simulations both volumes $L_s^D$ and
$L_l^D$ must be finite. The monotony criterion works equally well
in this case. Both volumes have to be sufficiently large in order to match
the conditions of the criterion
(which reflect the standard assumptions of finite size scaling).
Some care is needed to ensure that the susceptibility is actually measured at a
temperature that satisfies (\ref{monintro.45}), i.e. out of a region
where the difference in the volume dependence between first and
second order transitions appears.
In particular this concerns the lower bound on $\delta$ in
(\ref{monintro.45}) because for $L_l<\infty$,
also in case of a first order transition, a small neighbourhood around
the peak at $T_c(L)$ exists, where $\chi(T,L)$ is increasing in
volume because of the rounding of the $\delta$-singularity.

To summarize, let $\lambda$ denote a generic coupling constant.
Application of the monotony criterion amounts to a choice of two
sizes $L_s$ and $L_l$ with
$L_s$, $L_l$ and $L_l/L_s$ sufficiently large.
Let us define the ratio $r$
\be \label{ratio}
   r(L_s,L_l;\lambda) := 1 - \frac{ \chi(T_c(L_s)+\delta,L_s;\lambda)}
    {\chi(T_c(L_l)+\delta,L_l;\lambda)},
\ee
with $\delta$ satisfying the constraint (\ref{monintro.45}).
The monotony criterion then says that
\[
    r(L_s,L_l;\lambda) \;\; \left\{
    \begin{array}{r@{\; ,\;} l }
    > 0 & \mbox{2nd order} \\
    < 0 & \mbox{1st order }\\
    = 0 & \mbox{tricritical point for 
         $r$ changing sign at $\lambda$}.
    \end{array} \right. 
\]

In (\ref{ratio}) the susceptibility is measured at fixed distance to the
volume-dependent location of its maximum.
A reasonable alternative is to measure $\chi(T,L)$ for both volumes at
the same temperature $T$ regardless of $L$, but $T$ chosen as 
$T=T_c(\infty)+\delta$.
This requires that $T_c(L)$ approaches $T_c=T_c(\infty)$
according to
\be \label{shift}
  | T_c(L) - T_c(\infty) | \; \leq \; c \, \sigma(L),
\ee
cf.~(\ref{monintro.45}) and (\ref{running.peak}).
The behaviour (\ref{shift}) meets the standard finite size scaling
behaviour.

%
%




\begin{thebibliography}{9}
\bibitem{barber} M.~N.~Barber, in {\it Phase Transitions and Critical
    Phenomena}, Vol.~8, edited by C.~Domb and J.~L.~Lebowitz (Academic,
    New York), p.145
\bibitem{jsp} H.~Meyer-Ortmanns and T.~Reisz,
      J.~Stat.~Phys.~$\underline{87}$ (1997) 755-798

\end{thebibliography}
\end{document}